# The temperature dependence of the isothermal bulk modulus at 1 bar pressure

J. Garai[a)]

*Dept. of Earth Sciences, Florida International University, University Park, PC 344, Miami, FL 33199, USA*

A. Laugier

*IdPCES - Centre d'affaires PATTON, 6, rue Franz Heller, bât. A, F35700 Rennes, France*

It is well established that the product of the volume coefficient of thermal expansion and the bulk modulus is nearly constant at temperatures higher than the Debye temperature. Using this approximation allows predicting the values of the bulk modulus. The derived analytical solution for the temperature dependence of the isothermal bulk modulus has been applied to ten substances. The good correlations to the experiments indicate that the expression may be useful for substances for which bulk modulus data are lacking.

## I. INTRODUCTION

Any isothermal equation of state (EOS)[1-7] requires knowing the value of the bulk modulus at the temperature of interest. Theoretically the temperature dependence of the elastic constants can be determined as the sum of the anharmonic terms[8; 9]. At sufficiently low temperatures the elastic constant should vary as[10] $T^4$. Contrary to this suggestion some metallic substances have been found to show a $T^2$ rather than a $T^4$ dependence at low temperatures[11; 12]. There is no general prediction for higher temperatures. Experiments on refractory oxides, conducted at higher than room temperature, show a linear relationship between the bulk modulus and the temperature[13].

The third law of thermodynamics requires that the derivative of any elastic constant with respect to the temperature must approach zero as the temperature approaches absolute zero. Combining this criterion with the observed linear relationship at higher temperatures, Wachtman et al.[14] suggested an equation in the form of

$$B = B_0 - b_1 T e^{\left(-\frac{T_0}{T}\right)}, \tag{1}$$

where $B_0$ is the bulk modulus at absolute zero, T is the temperature, and $b_1$ and $T_0$ are arbitrary constants. Theoretical justification for Wachtman's Equation was suggested by Anderson[15].

Based on shock-wave and static-compression measurements on metals, a linear relationship between the logarithm of the bulk modulus and the specific volume has been detected for metals[16]:

$$\ln B_T = \ln B_0 + \alpha\left(\frac{\Delta V}{V}\right), \quad (2)$$

where $\alpha$ is a constant depending on the material. The linear correlation is valid up to 40% volume change. Using this linear correlation Jacobs and Oonk[17] proposed a new equation of state. They rewrite equation (2) as:

$$V_m^0(T) = V_m^0(T_0) + b\ \ln\left(\frac{B^0(T)}{B^0(T_0)}\right), \quad (3)$$

where $V_m^0$ denotes molar volume, $T_0$ the reference temperature and the superscript "0" refers to standard pressure (1 bar). Equation (3) successfully reproduces the available experimental data[17-19] for MgO, $Mg_2SiO_4$, and $Fe_2SiO_4$.

In this study, analytical solution for the temperature dependence of the bulk modulus and the pressure dependence of the volume coefficient of thermal expansion has been derived from fundamental thermodynamic equations. The derived theoretical expression for the temperature dependence of the isothermal bulk modulus is compared to experiments.

## II. THEORY

The isothermal bulk modulus $[B_T]$ is defined as:

$$B_T \equiv -V\left(\frac{\partial p}{\partial V}\right)_T. \quad (4)$$

It is assumed that the solid is homogeneous, isotropic, non-viscous and has linear elasticity. It is also assumed that the stresses are isotropic; therefore, the principal stresses can be identified as the pressure[20] $p = \sigma_1 = \sigma_2 = \sigma_3$.

The definition for the volume coefficient of thermal expansion ($\alpha_{Vp}$) is given as:

$$\alpha_{Vp} \equiv \frac{1}{V}\left(\frac{\partial V}{\partial T}\right)_p. \quad (5)$$

Both the volume coefficient of expansion and the isothermal bulk modulus are pressure and temperature dependent; therefore, the universal description of solids requires knowing the derivatives of these parameters.

$$\left(\frac{\partial \alpha_V}{\partial T}\right)_p \; ; \; \left(\frac{\partial \alpha_V}{\partial p}\right)_T \; ; \; \left(\frac{\partial B_T}{\partial T}\right)_p \; ; \; \left(\frac{\partial B_T}{\partial p}\right)_T \tag{6}$$

The complete thermo-physical description of an elastic solid requires knowing the two parameters [Eq. (4) and (5)] and their four derivatives [Eq. (6)].

The pressure derivative of the volume coefficient of thermal expansion and the temperature derivative of the isothermal bulk modulus are not independent from each other and the relationship between these derivatives is given[5]:

$$\left(\frac{\partial \alpha_{V_p}}{\partial p}\right)_T = \frac{1}{B_T^2}\left(\frac{\partial B_T}{\partial T}\right)_p. \tag{7}$$

It is worth to note that Eq. (7) reduces the number of independent partial derivatives of the fundamental parameters $\alpha_{V_p}$ and $B_T$ to three.

The definition of the isothermal Anderson-Grüneisen parameter [$\delta_T$] is

$$\delta_T = -\frac{1}{\alpha_{V_p} B_T}\left(\frac{\partial B_T}{\partial T}\right)_p = -\frac{1}{\alpha_{V_p}}\left(\frac{\partial \ln B_T}{\partial T}\right)_p. \tag{8}$$

Eq. (8) can be written as:

$$B_T = B_{T=0} e^{-\int_{T=0}^{T} \alpha_V(p,T)\delta(p,T)dT} \tag{9}$$

At one bar pressure Eq. (9) reduces to:

$$B_T^\circ = B_{T=0}^\circ e^{\int_{T=0}^{T} \alpha_V^\circ(T)\delta^\circ(T)dT}, \tag{10}$$

where subscript [°] refers to 1 bar pressure. Inserting Eq. (7) into Eq. (8) gives:

$$\delta_T = -\frac{1}{\alpha_V B_T}\left\{B_T^2\left(\frac{\partial \alpha_V}{\partial p}\right)_T\right\} = -\frac{B_T}{\alpha_V}\left(\frac{\partial \alpha_V}{\partial p}\right)_T = -B_T\left(\frac{\partial \ln \alpha_V}{\partial p}\right). \tag{11}$$

Eq. (11) can also be written as:

$$\alpha_{V_p} = \alpha_{V_{p_o}} e^{-\int_{p^o}^{p} \frac{\delta(p,T)}{B(p,T)}dp}, \tag{12}$$

where p° denotes 1 bar pressure. An alternative derivation of Eq. (10) and (12) is given in the appendix.

Relations Eq. (10) and (12) are generally valid. The difficulty is that the Anderson-Grüneisen parameter in these equations is not constant but rather changes with temperature, especially at low temperatures. Inspecting Eq. (8) reveals that it is composed of the thermal pressure derivative with respect to temperature:

$$\left(\frac{\partial p^{th}}{\partial T}\right)_V = \alpha_{V_p} B_T, \qquad (13)$$

and the temperature derivative of the isothermal bulk modulus.

For substances which the quasi-harmonic approximation (QHA) model is applicable, the product of the volume coefficient of thermal expansion and isothermal bulk modulus is nearly constant above the Debye temperature[21]. The constant value for the product of the volume coefficient of thermal expansion and the isothermal bulk modulus at temperatures higher than the Debye temperature is also consistent with experiments[22-25]. Inspecting Figure 1 shows that the temperature derivative of the isothermal bulk modulus is nearly constant at temperatures above the Debye temperature. Therefore it is reasonable to assume that the Anderson- Grüneisen parameter is approximately constant at temperatures above the Debye temperature.

Assuming that $\alpha_{V_p} B_T$ is constant at temperatures higher than the Debye temperature then three parameters, the volume coefficient of thermal expansion, the bulk modulus, and the Anderson-Grüneisen parameter can completely describe the relationship between the pressure, volume and temperature. In this study the validity of Eq. (10) will be investigated by comparing the theoretically derived expression to experiments.

## III. RESULTS AND DISCUSSION

Experiments with ten or more data point were chosen from the literature in order to evaluate the theoretically derived temperature dependence of the bulk modulus. Experiments of Ag, Au, MgO, $Al_2O_3$, $MgAl_2O_4$, $Mg_2SiO_4$, $(Fe_{0.1}Mg_{0.9})_2SiO_4$, CaO, NaCl, and KCl were used for the investigation.

The integration of $\int_{T=0}^{T} \alpha_{V_p} dT$ was done numerically by using linear polynomials. The volume coefficient of thermal expansion values at various temperatures were taken from ref. 26 for Ag and Au, and from ref. 21 for the rest of the substances. The polynomial fit of seven experiments given by Jacobs and Oonk[27]:

$$\alpha_{V_p}(p_0, T) = 4.5248 \times 10^{-5} + 8.4711 \times 10^{-10} T - 4.1959 \times 10^{-3} T^{-1} + 2.4984 \times 10^{-12} T^2 \qquad (14)$$

was also used to determine the values of the volume coefficient of thermal expansion for MgO. If experiments are not available then the linear correlation between the heat capacity and the volume coefficient of thermal expansion can be used calculating the value of the volume coefficient of thermal expansion at the temperature of interest[28]. Using least-squares fit the Anderson-Grüneisen parameter and the correlation coefficients were determined. The calculated Anderson-Grüneisen parameters are in very good agreement with experiments. The correlation coefficients are the lowest for the two noble metals 0.9972 and 0.9987, while for the rest of the minerals the values are between 0.9992 and 1.0. At low temperatures the Anderson-Grüneisen parameter changes significantly as a function of temperature[29], and the constant value approach for the Anderson-Grüneisen parameter might not be appropriate. The data of the noble metals contains the very low temperature experiments, which can explain the slightly weaker correlation.

The best fits for the two different data sets of MgO resulted almost identical values for the isothermal bulk modulus at zero temperature and the Anderson-Grüneisen parameters. The results are given in Table 1. The calculated values were plotted against experiments (Fig. 1). It can be seen that the derived theoretical relationship for the temperature dependence of the bulk modulus can reproduce the experimental values with high accuracy for the entire temperature range of the solid phase.

## IV. CONCLUSIONS

Traditionally the bulk modulus has to be measured at the temperature of interest requiring numerous experiments. Assuming constant value for the product of the isothermal bulk modulus and the volume coefficient of thermal expansion allows describing the volume-pressure-temperature relationship of a solid from three parameters, namely from the zero temperature values of the volume coefficient of thermal expansion, and isothermal bulk modulus, and from Anderson-Grüneisen parameter. The derived theoretical expression, Eq.(10), can be employed to extrapolate data measured at convenient temperatures to the temperature of interest.

Our investigation showed that assuming constant value for the Anderson-Grüneisen parameter at temperatures higher than the Debye temperature is reasonable.

**ACKNOWLEDGEMENT:** We would like thank the anonymous reviewer for the constructive comments and suggestions, which helped to improve the quality of the manuscript.

**Appendix 1.**

The temperature dependence of the bulk modulus and the pressure dependence of the volume coefficient of thermal expansion can also be derived from fundamental thermodynamic relationships, which do not include the definition of the Anderson-Grüneisen parameter. Using the Euler's chain relation

$$-1 = \left(\frac{\partial p}{\partial V}\right)_T \left(\frac{\partial V}{\partial T}\right)_p \left(\frac{\partial T}{\partial p}\right)_V = -\frac{B_T}{V} \times V\alpha_{V_p} \times \left(\frac{\partial T}{\partial p}\right)_V. \quad (15)$$

gives the pressure and temperature relationship at constant volume as:

$$B_T \alpha_{V_p} = \left(\frac{\partial p}{\partial T}\right)_V. \quad (16)$$

Combining Eq. (7) and (16) gives

$$\left(\frac{\partial \alpha_{V_p}}{\partial p}\right)_T = \frac{\alpha_{V_p}}{B_T} \left(\frac{\partial T}{\partial p}\right)_V \left(\frac{\partial B_T}{\partial T}\right)_p. \quad (17)$$

Introducing a parameter (a), which is defined as:

$$a = \frac{1}{\alpha_{V_p}^2} \left(\frac{\partial \alpha_{V_p}}{\partial p}\right)_T \left(\frac{\partial p}{\partial T}\right)_V = -\left[\frac{\partial}{\partial p}\left(\frac{1}{\alpha_{V_p}}\right)\right]_T \left(\frac{\partial p}{\partial T}\right)_V. \quad (18)$$

and substituting this parameter into Eq. (17) gives

$$d\ln B_T = a\alpha_{V_p} dT. \quad (19)$$

The integral of Eq. (19) gives the temperature dependence of the bulk modulus as:

$$B_T = B_{T=0} e^{\int_{T=0}^{T} a\alpha_{V_p} dT}. \quad (20)$$

The pressure dependence of the volume coefficient of thermal expansion can also be determined from Eq. (17). Rearranging Eq. (17) gives

$$d\ln \alpha_{V_p} = \left(\frac{\partial T}{\partial p}\right)_V \left(\frac{\partial B_T}{\partial T}\right)_p \frac{dp}{B_T}. \quad (21)$$

Defining a dimensionless parameter b as:

$$b = \left(\frac{\partial T}{\partial p}\right)_V \left(\frac{\partial B_T}{\partial T}\right)_p \quad (22)$$

and integrating Eq. (21) leads to

$$\alpha_{V_p} = \alpha_{V_{p=0}} e^{\int_{p=0}^{p} \frac{b}{B_T} dp}. \quad (23)$$

Eq. (23) describes the pressure dependence of the volume coefficient of thermal expansion.

Introducing the symbol $p'_{th}$ for the partial derivative of the thermal pressure with respect to the temperature

$$p'_{th} = \left(\frac{\partial p}{\partial T}\right)_V = \alpha_{V_p} B_T \quad \Leftrightarrow \quad \frac{1}{B_T} = \frac{\alpha_{V_p}}{p'_{th}} \tag{24}$$

the parameters a and b can be defined as:

$$a = \frac{p'_{th}}{\alpha_{V_p}^2}\left(\frac{\partial \alpha_V}{\partial p}\right)_T, \text{ and } b = \frac{1}{p'_{th}}\left(\frac{\partial B_T}{\partial T}\right)_p = -\delta_T \tag{25}$$

where $\delta_T$ is the isothermal Anderson-Grüneisen parameter given by:

$$\delta_T = -\frac{1}{\alpha_{V_p} B_T}\left(\frac{\partial B_T}{\partial T}\right)_p = -\frac{1}{\alpha_p}\left(\frac{\partial \ln B_T}{\partial T}\right)_p. \tag{26}$$

Using the definition of the coefficient a [Eq. (25)], and combining with Eq. (7), and the expression of the temperature derivative of the thermal pressure Eq. (24) implies that the two parameters a and b are equal with each other.

$$a = \frac{\alpha_{V_p} B_T}{\alpha_{V_p}^2}\frac{1}{B_T^2}\left(\frac{\partial B_T}{\partial T}\right)_p = \frac{1}{p'_{th}}\left(\frac{\partial B_T}{\partial T}\right)_p = b = -\delta_T \tag{27}$$

where b was substituted by using Eq. (25). The identity (a = b) holds true regards of the temperature. The temperature dependence of the bulk modulus then can be written as:

$$B_T = B_{T=0} e^{-\int_{T=0}^{T} \delta_T \alpha_{V_p} dT} \tag{28}$$

while the pressure dependence of the volume coefficient of thermal expansion

$$\alpha_{V_p} = \alpha_{V_{p=0}} e^{-\int_{p=0}^{p} \frac{\delta_T}{B_T} dp}. \tag{29}$$

Equations (28) and (29) recover equations (10) and (12) respectively.

TABLE I.  $B_0$ is the bulk modulus at zero temperature, $\delta_{average}^{calculated}$ is the calculated average of Anderson-Grüneisen parameter in the temperature range of interest, $\delta_T$ is the Anderson-Grüneisen parameter, R is the correlation coefficient, and N is the number of experiments. The data for Ag and Au is from ref. 29-33 while the rest of the data is from ref. 21. The volume coefficient of thermal expansion values for MgO(2) are from ref. 27. The errors represent the standard deviation.

| Material | $B_0$ [GPa] | $\delta_{average}^{calculated}$ | $\delta_T$ | R | N |
|---|---|---|---|---|---|
| Ag | 165.2(9)[30] | 5.22(86) | 5.66[31]; 6.18[32] | 0.9972 | 14 |
| Au | 230.8(6)[30] | 5.17(56) | 5.21-6.39[29; 33] | 0.9987 | 14 |
| MgO | 165.4(6) | 4.91(8) | 4.66-5.26 | 0.9998 | 18 |
| MgO(2) | 165.3(5) | 4.93(8) | 4.66-5.26 | 0.9998 | 18 |
| $Al_2O_3$ | 255.6(8) | 5.21(12) | 4.50-5.71 | 0.9997 | 16 |
| $MgAl_2O_4$ | 212.0(7) | 6.75(27) | 6.24-7.73 | 0.9993 | 15 |
| $Mg_2SiO_4$ | 130.1(2) | 5.46(4) | 5.42-5.94 | 0.99997 | 15 |
| $(Fe_{0.1}Mg_{0.9})_2SiO_4$ | 133.8(9) | 5.44(25) | 5.43-6.59 | 0.9992 | 13 |
| CaO | 116.0(3) | 5.06(9) | 5.01-6.19 | 0.9999 | 10 |
| NaCl | 26.8(2) | 5.96(16) | 5.56-6.53 | 0.9997 | 10 |
| KCl | 18.8(2) | 5.92(21) | 5.84-6.19 | 0.9996 | 12 |

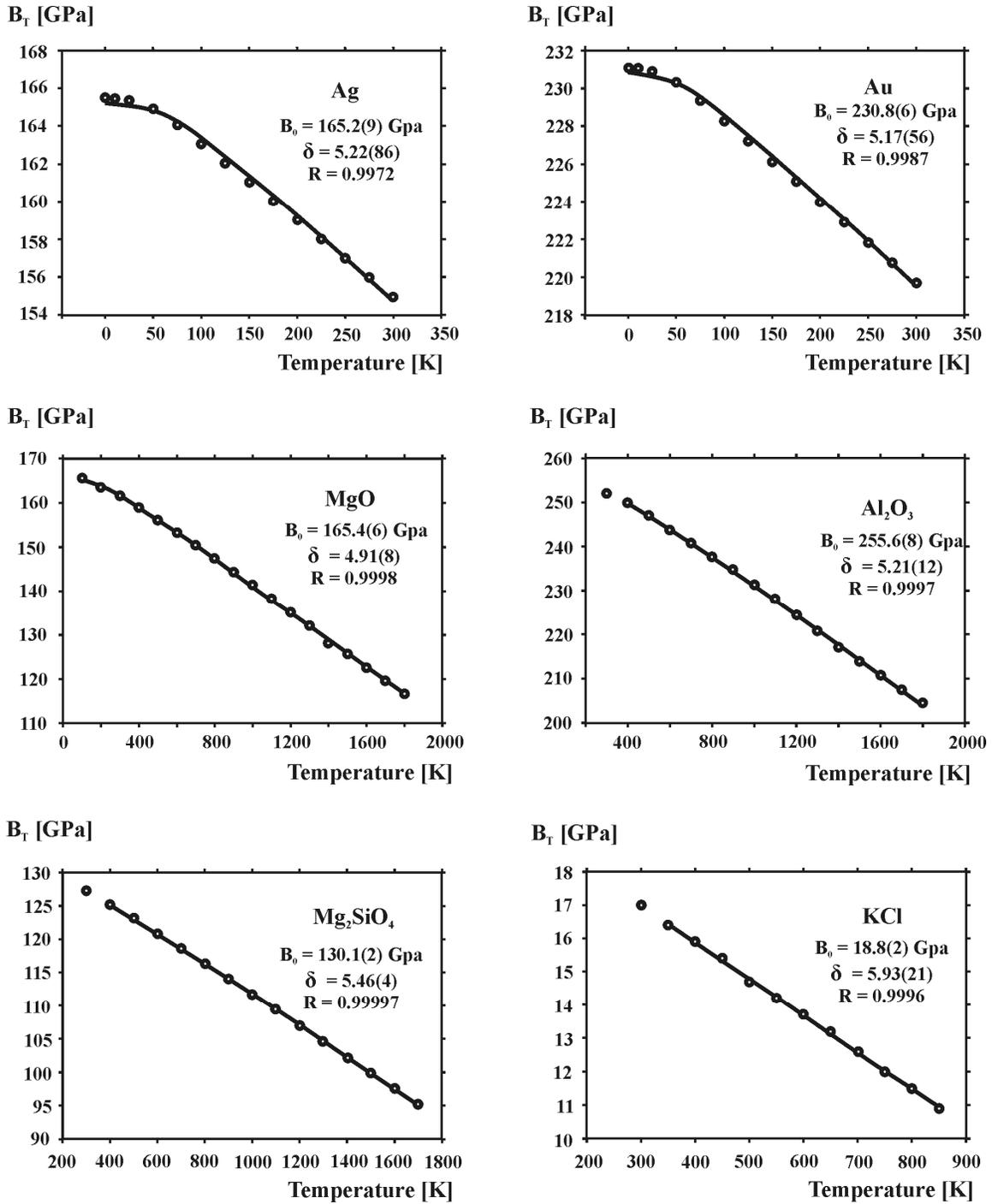

FIG. 1 The solid lines are the calculated bulk modulus using Eq.(10), while the dots represent the experimental values. $B_0$ is the bulk modulus at zero temperature, $\delta$ is the calculated average Anderson-Grüneisen parameter, and R is the correlation coefficient.